\documentclass[superscriptaddress,reprint, nofootinbib,
 amsmath,amssymb,aps]{revtex4-1}

\usepackage{graphicx}
\usepackage{gensymb}
\usepackage{xcolor}
\usepackage{subcaption}
\usepackage{graphicx}
\usepackage{dcolumn}
\usepackage{bm}
\usepackage{float}
\usepackage[percent]{overpic}
\usepackage[%
  colorlinks=true,
  urlcolor=blue,
  linkcolor=blue,
  citecolor=blue
]{hyperref}
\usepackage{makecell}

\begin{document}

\title{Predicted Complex Lithium Phases at Terapascal Pressures}

\author{Jack Whaley-Baldwin}
\email{jajw4@cam.ac.uk}
\affiliation{TCM Group, Cavendish Laboratory, University of Cambridge}

\author{Miguel Martinez-Canales}
\affiliation{Scottish Universities Physics Alliance (SUPA), School of Physics and Astronomy and Centre for Science at Extreme Conditions, University of Edinburgh, Edinburgh EH9 3FD, UK}

\author{Chris J. Pickard}
\affiliation{Department of Materials Science and Metallurgy, University of Cambridge, 27 Charles Babbage Road,
Cambridge CB3 0FS, United Kingdom}
\affiliation{Advanced Institute for Materials Research, Tohoku University, 2-1-1 Katahira, Aoba, Sendai 980-8577, Japan}

\date{\today}

\begin{abstract}
We investigate the pressure-temperature ($p$-$T$) phase diagram of elemental lithium (Li) up to multiterapascal (TPa) pressures using \textit{ab-initio} random structure search (AIRSS) and density functional theory (DFT). At zero temperature, beyond the high-pressure $Fd\bar{3}m$ diamond structure predicted in previous studies, we find eleven solid-state phase transitions to structures of greatly varying complexity, in addition to two structures that we calculate will become stable with sufficient temperature. The full $p$-$T$ dependence of the phase boundaries are computed within the vibrational harmonic approximation, and the solid-liquid melting line is calculated using \textit{ab-initio} molecular dynamics simulations. Notably, between $39.1$ TPa and $55.7$ TPa, Li adopts an elaborate monoclinic structure with 46 atoms in the primitive unit cell, and between $71.9$ TPa and $103$ TPa, an incommensurate host-guest phase of the Ba-IV type. We find that Li, hitherto predicted to be an electride at TPa pressures, abruptly loses its electride character above $16$ TPa, reverting back to normal metallic behaviour with a corresponding rise in the Fermi-level electronic density of states (eDOS) and broadening of the electronic bands.
\end{abstract}

\maketitle

\section{\label{intro}Introduction}

Under compression, lithium (Li) transforms through a multitude of crystal structures with varying structural complexity and markedly different electronic properties, which in-turn exhibit metallic \cite{Shimizu_2002}, superconducting \cite{Shimizu_2002} and semiconducting \cite{Marques_2011,Lv_2011} behaviour. At zero temperature and pressure, the ground state structure of Li is face-centered cubic (fcc) \cite{isotope_effects_lithium}, and it is a simple metal well-described by the nearly-free electron (NFE) model. With an increase in pressure, experiment has shown that Li transitions to the hR1 phase (space group
$R\bar{3}m$) at 39 GPa, and the cI16 phase (space group $I\bar{4}3d$) at $42$ GPa \cite{Hanfland_lithium_2000,cold_melting_lithium,Frost_2019}. Further transitions to the exceptionally complex oC88, oC40 and oC24 structures occur at $61$ GPa, $70$ GPa and $115$ GPa respectively \cite{Marques_2011,cold_melting_lithium,Lv_2011,Frost_2019}, with oC24 being a narrow- or zero- gap semiconductor \cite{Marques_2011,Lv_2011,semiconducting_lithium}, and oC40 being, remarkably, a wide-gap semiconductor \cite{Marques_2011,Lv_2011}. In addition to exhibiting a rich solid-state phase transition sequence, the melting line of Li also displays intriguing behaviour with increasing pressure; namely, a pronounced drop in melting temperature at $50$ GPa to just $190$ K, implying that Li can be brought into a liquid state upon compression even at temperatures well below room temperature (`cold melting') \cite{cold_melting_lithium,hernandez_lithium,electride_lithium,Frost_2019}.
\par
First-principles computational structure searches have further extended the high-pressure phase diagram of Li, with predicted transitions to a $P4_{2}/mbc$ structure at $247$ GPa, $R\bar{3}m$ structure at $449$ GPa, and eventually the $Fd\bar{3}m$ diamond structure ($483$ GPa) \cite{Lv_2011,pickard_needs_2009}.
\par
Whilst the very highest static laboratory pressures are typically only several hundred GPa, dynamic compression techniques have allowed pressures of several terapascal (TPa) to be accessed \cite{ramped_compression_2TPa,gold_and_platinum_standards,copper_experiment_terapascal,tin_experiment_terapascal,molybdenum_experiment_terapascal,gold_experiment_terapascal,silicon_experiment_terapascal,jeanloz_terapascal_2007,bridgmanite_compression,ramp_compression_magnesium_oxide,lithium_fluoride_terapascal,iron_exoplanets,Gorman_2022}, with one exceptional experiment exceeding $5$ TPa \cite{ramped_compression_5TPa_diamond}. Concomitant with this progress in TPa-pressure generation, advances in x-ray diffraction techniques have allowed the crystal structures of such TPa-compressed matter to be calculated \cite{powder_diffraction_terapascal,xray_diffraction_NIF,Coppari2021}, and there is a particular interest in determining the structure of pure elements under pressure, especially in using their equation of state (EoS) as a reference standard \cite{ramped_compression_5TPa_diamond,ramped_compression_2TPa,gold_and_platinum_standards,copper_experiment_terapascal,tin_experiment_terapascal,molybdenum_experiment_terapascal,gold_experiment_terapascal,silicon_experiment_terapascal}. In determining such EoSs, computational structure search provides invaluable insight which complements the experimental results, and is especially important when the target system is difficult to access experimentally. Li itself is such an example; having just three electrons, it is a difficult diffraction target and possesses an extremely small x-ray absorption cross-section, and its two main isotopes show neutron absorption cross-sections that differ by several orders of magnitude \cite{li_neutron_diffraction_1,li_neutron_diffraction_2,li_neutron_diffraction_3}. Li is also highly reactive with materials commonly used to achieve multi-TPa pressures \cite{Hanfland_lithium_1999}, and therefore a computational search may, for the time being, provide the only insight into its behaviour at TPa pressures.

\begin{figure*}[ht!]
    \centering
    
    \begin{subfigure}{0.490\textwidth}
        \centering
        \includegraphics[width=0.995\textwidth]{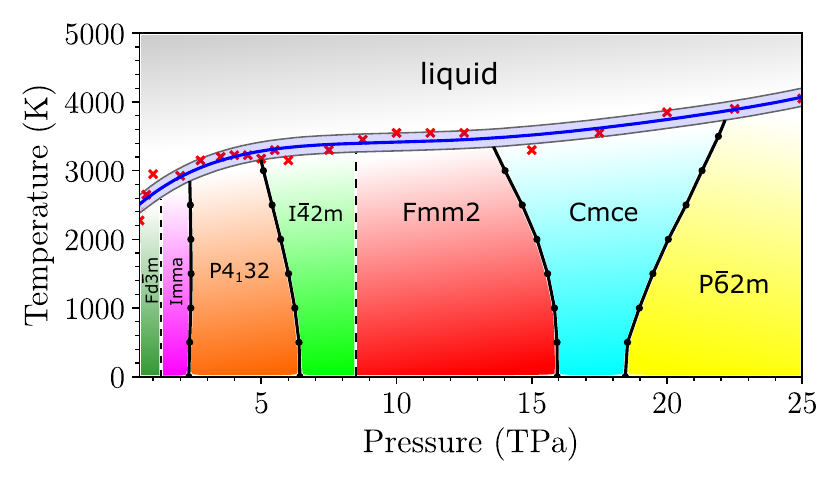}
        \label{LOW_finite_T}
    \end{subfigure}
    \hspace{0.15cm}
    \begin{subfigure}{0.490\textwidth}
        \centering
        \includegraphics[width=0.995\textwidth]{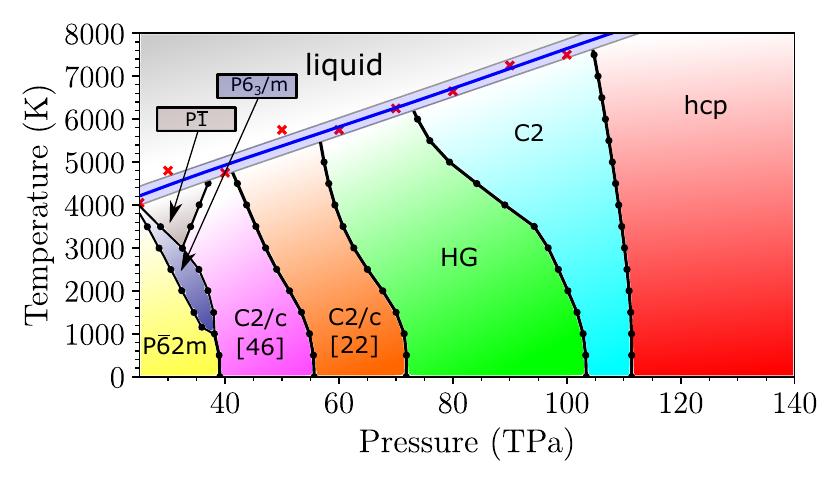}
        \label{HIGH_finite_T}
    \end{subfigure}
    
    \vspace{-0.4cm}
    
    \caption{Phase diagram of elemental Li up to $140$ TPa. Each coloured region denotes the structure with the lowest Gibbs free energy, with $G(p,T) = U_{elec} + U_{phonon} - TS_{elec} - TS_{phonon} + pV$. Solid-state vibrational effects have been included at the level of the harmonic approximation (HA), and solid circles on the phase boundaries denote explicitly calculated $(p,T)$ points according to the HA. The melting line, as calculated using \textit{ab-initio} molecular dynamics simulations (red crosses), is  fitted to a fifth-order polynomial and is shown as a solid blue line, with a thickness according to its uncertainty. Dashed vertical lines indicate second-order soft-phonon transitions.}
    \label{finite_T_diagrams}
\end{figure*}

As a high-pressure element of interest, Li is an extremely important case that lies between the quantum nuclear effect dominated hydrogen \cite{Pickard_hydrogen_2012,Drummond_hydrogen_2015} and helium \cite{Monserrat_helium_2014} systems, and the rest of the periodic table. In the TPa regime, computational structure searches have been performed on neighbouring group-I elements sodium (Na) \cite{Pickard_sodium_2015} and potassium (K) \cite{Whaley-Baldwin_potassium_2023}, as well as other first-row elements carbon (C) \cite{Martinez_carbon_2012}, Nitrogen (N) \cite{Sun_nitrogen_2013}, Oxygen (O) \cite{Sun_oxygen_2012} and Fluorine (F) \cite{Duan_fluorine_2021}, but we are unaware of any such work on Li. In addition to laying the groundwork for any future TPa-experimental work on Li, a first-principles search in Li is also of great interest to determine whether the remarkable structural complexity of Li persists at multi-TPa pressures.
\par
In this work, we extend the pressure-temperature ($p$-$T$) phase diagram of Li up to multiterapascal pressures, proposing eleven solid-state phase transitions beyond the diamond $Fd\bar{3}m$ phase discovered in previous searches \cite{pickard_needs_2009}, in addition to two structures that are stabilized at higher temperatures. We further calculate the full $p$-$T$ dependence of the solid-liquid melting line using \textit{ab-initio} molecular dynamics simulations, and provide a detailed analysis of the nature of bonding in Li through extensive calculations of the electronic charge density at different pressures.

\section{\label{comp_details}Computational Details}
\vspace{-0.2cm}
We used the \textit{ab-initio} Random Structure Searching (AIRSS) package \cite{airss_PRL,airss_JPCM} to perform our structure search, with the first-principles Density Functional Theory (DFT) code \verb|CASTEP| \cite{castep} performing the underlying electronic structure calculations. We performed structure searches at $1$, $2.5$, $5$, $7.5$, $10$, $15$, $25$, $50$, $100$ and $150$ TPa, using randomly constructed unit cells containing between 1 and 64 atoms, and possessing between 1 and 48 randomly chosen symmetry operations. Each search generated approximately $6,000$ relaxed structures, giving $\sim 60,000$ structures in total.  Due to the appearance of a large number of complex structures around $50$ TPa, we further constructed a machine-learned potential specifically at this pressure using the Ephemeral Data Derived Potential (EDDP) package \cite{eddp}, which allowed us to search in larger cells containing up to 256 atoms, and which generated $\sim 500,000$ structures. For both the structure searches and our final results presented here, we constructed two all-electron ultrasoft pseudopotentials with cutoff radii of $0.40$ Bohr (for calculations at $100$ TPa and above) and $0.60$ Bohr (for all other calculations), requiring plane-wave cutoffs of $4.50$ keV and $2.25$ keV respectively to converge relative energies to better than $0.1$ meV per atom. We used Monkhorst-Pack \textbf{k}-point grids with a sampling spacing of $2\pi \times 0.025$ \AA$^{-1}$. Full details of our calculations can be found in the Supplemental Material \cite{supp}.

\section{\label{results_and_discussion}Results and Discussion}

The $p$-$T$ phase diagram of Li, as calculated using the PBE exchange-correlation functional \cite{PBE} and with ionic vibrations included at the level of the harmonic approximation, is shown in Fig \ref{finite_T_diagrams}. We find that vibrational contributions to the free energy are very important in Li, observing a substantial variance in the vibrational energies of the different structures - much larger than the static lattice enthalpy differences between these structures - which are sufficient to substantially change the phase transition sequence when compared with the static-lattice picture. We note that similar observations have been made in Li before \cite{isotope_effects_lithium,Ashcroft_1989}, and it is known that the neglect of vibrational effects in Li leads to an incorrect phase diagram \cite{Gorelli_lithium_2012}. Several structures that are the lowest in enthalpy at the static lattice level become unfavourable when vibrations are included. Below $1.30$ TPa, at zero temperature, we find that Li adopts the diamond structure ($Fd\bar{3}m$ symmetry), in agreement with previous works \cite{Lv_2011,pickard_needs_2009}. At $1.30$ TPa, Li undergoes a symmetry-lowering distortion via a soft optical phonon mode at the $\Gamma$-point to a structure of $Imma$ symmetry with 2 atoms in the primitive cell. We then find further transitions to structures of gradually increasing complexity, with phases of $P4_{1}32$, $I\bar{4}2m$, $Fmm2$, $Cmce$ and $P\bar{6}2m$ symmetry at $2.32$ TPa, $6.43$ TPa, $8.50$ TPa, $15.9$ TPa and $18.5$ TPa respectively. The $P4_{1}32$ structure has 4 atoms in its primitive cell, the $I\bar{4}2m$ and $Fmm2$ structures have 10 atoms, and the $Cmce$ and $P\bar{6}2m$ phases possess 14 atoms in their primitive cells.
\par
At $39.1$ TPa, Li undergoes a transition to a remarkably complex structure of $C2/c$ symmetry, with 46 (92) atoms in its primitive (conventional) cell. This structure (denoted as $C2/c \ [46]$) comprises interpenetrating networks of Li atoms, similar to phases adopted by Li at much lower pressures of several hundred GPa \cite{Lv_2011,pickard_needs_2009}. In fact, we find that several highly complex polymorphs become very close to being lowest in enthalpy around 50 TPa, and two of these structures, 22-atom $P6_{3}/m$ and 43-atom $P\bar{1}$, possess sufficient vibrational entropy to become stabilised at finite temperatures within relatively small regions of the $p$-$T$ plane; above ~1500 K and in the pressure range 25-37 TPa (see Fig \ref{finite_T_diagrams}). Given the appearance of these complex structures, we constructed a machine-learned potential at $50$ TPa using the Ephemeral Data Derived Potential (EDDP) package \cite{eddp}. The use of a machine-learned potential allowed us to perform structure searches in much larger cells and with a computational expense orders of magnitude smaller than the equivalent DFT calculations. The resulting search generated over half a million structures with up to 256 atoms in the unit cell, but none were found to be lower in free energy than the 46-atom $C2/c$ in any part of the $p$-$T$ plane. We note that Wang \textit{et. al.} also used a machine-learned potential to investigate highly complex structures of Li, albeit at much lower pressures \cite{ma_mlp}. Similar to our work here, the authors found many competitive structures but were unable to find any $p$-$T$ region in which the new structures became lowest in free energy. Eventually, Li transitions from $C2/c \ [46]$ to a 22-atom $C2/c$ structure ($C2/c \ [22]$) at $55.7$ TPa.

\addtolength{\tabcolsep}{+2.5pt} 
\renewcommand{\arraystretch}{1.25}
\begin{table}
\begin{tabular}{c c c}
\hline\hline
\centering
Structure    & Pressure (TPa) & Interstitial Charges ($|e|$) \\  \hline
$Imma$       & 2.00           & $-0.456$ (4e)                \\  \hline
$P4_{1}32$   & 5.00           & $-0.100$ (8c)                \\  \hline
$I\bar{4}2m$ & 7.50           & $-0.443$ (4c)                \\
             &                & $-0.292$ (8i)                \\  \hline
$Fmm2$       & 12.0           & $-0.419$ (8b)                \\
             &                & $-0.283$ (8c)                \\
\hline\hline
\end{tabular}
\captionof{table}{Values and locations of the interstitial electronic charges for the electride phases discovered in this work, as obtained from a Bader charge analysis. Wyckoff positions are shown in brackets.}
\label{bader_table}
\addtolength{\tabcolsep}{2.5pt} 
\end{table}

\begin{figure}[h!]
    \vspace{0.20cm}
    \centering
    \hspace{-0.6cm}
    \includegraphics[width=0.475\textwidth]{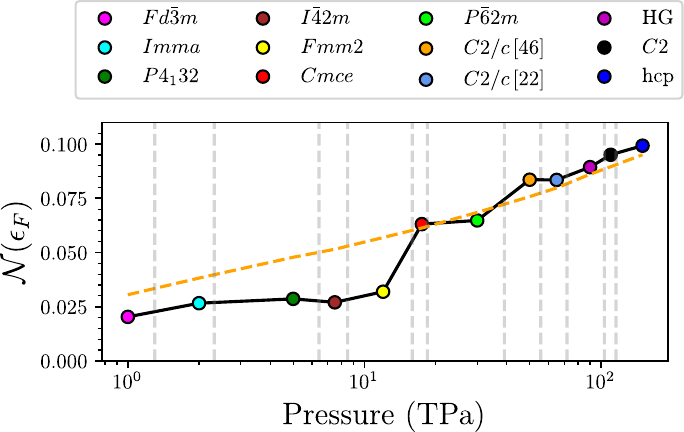}
    \caption{Fermi-level electronic density of states (eDOS) in units of states eV$^{-1}$ \AA $^{-3}$ for the phases discussed in this work. The grey vertical dashed lines mark the structural phase transitions, and the dashed orange line is the corresponding free-electron eDOS.}
    \label{edos_freqs_etc}
\end{figure}

\begin{figure}[h!]
    \centering

    \begin{subfigure}{0.425\textwidth}
        \centering
        \hspace{-1cm}
        \includegraphics[width=0.950\textwidth]{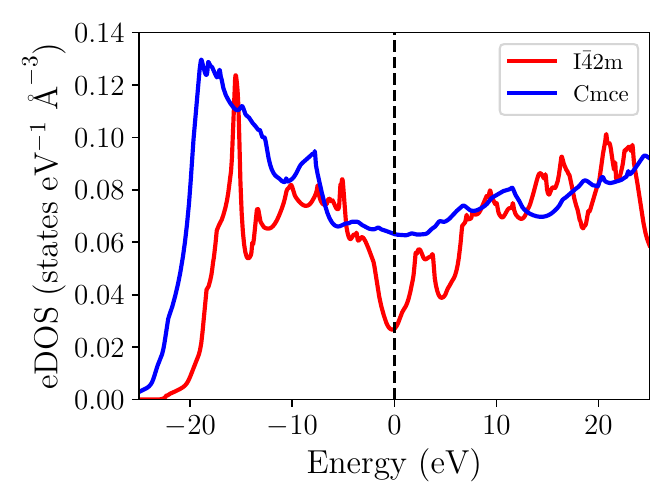}
        \caption{Electronic density of states (eDOS) for the $I\bar{4}2m$ ($7.50$ TPa) and $Cmce$ ($17.5$ TPa) structures.}
        \label{edos}
    \end{subfigure}

    \vspace{0.25cm}
    
    \begin{subfigure}{0.425\textwidth}
        \centering
        \hspace{-1cm}
        \includegraphics[width=0.950\textwidth]{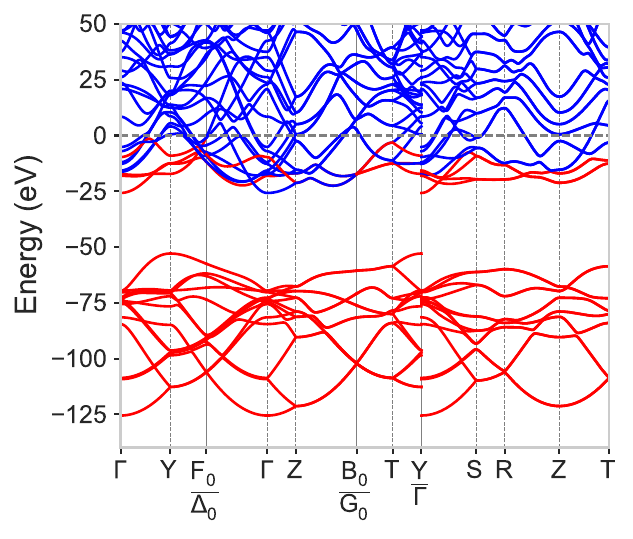}
        \caption{$Cmce$ bandstructure at $17.5$ TPa.}
        \label{Cmce_BS}
    \end{subfigure}
    
    \vspace{0.5cm}
    
    \begin{subfigure}{0.425\textwidth}
        \centering
        \hspace{-1cm}
        \includegraphics[width=0.950\textwidth]{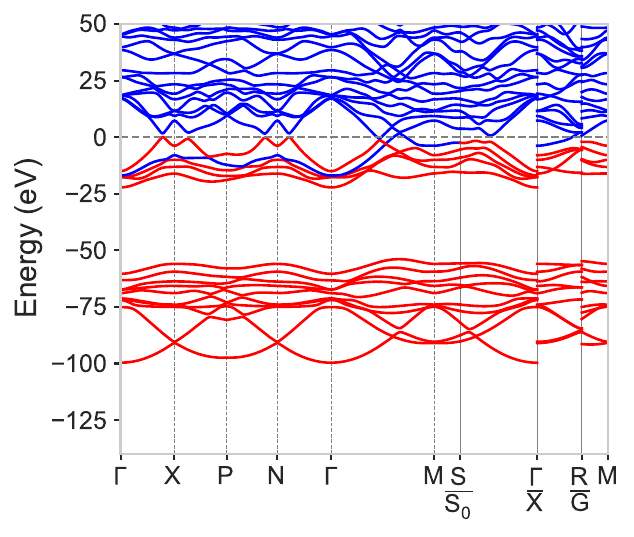}
        \caption{$I\bar{4}2m$ bandstructure at $7.50$ TPa.}
        \label{I-42m_BS}
    \end{subfigure}
    
    \caption{Electronic DOS and bandstructures for the $Cmce$ (non-electride) and $I\bar{4}2m$ (electride) structures. In Figs. \ref{Cmce_BS} and \ref{I-42m_BS}, the bands are coloured red (blue) if they are below (above or cross) the Fermi level.}
    \label{edos_and_bandstructures}
\end{figure}

\begin{figure}[h!]
    \centering
    \hspace{-1cm}
    \includegraphics[width=0.300\textwidth]{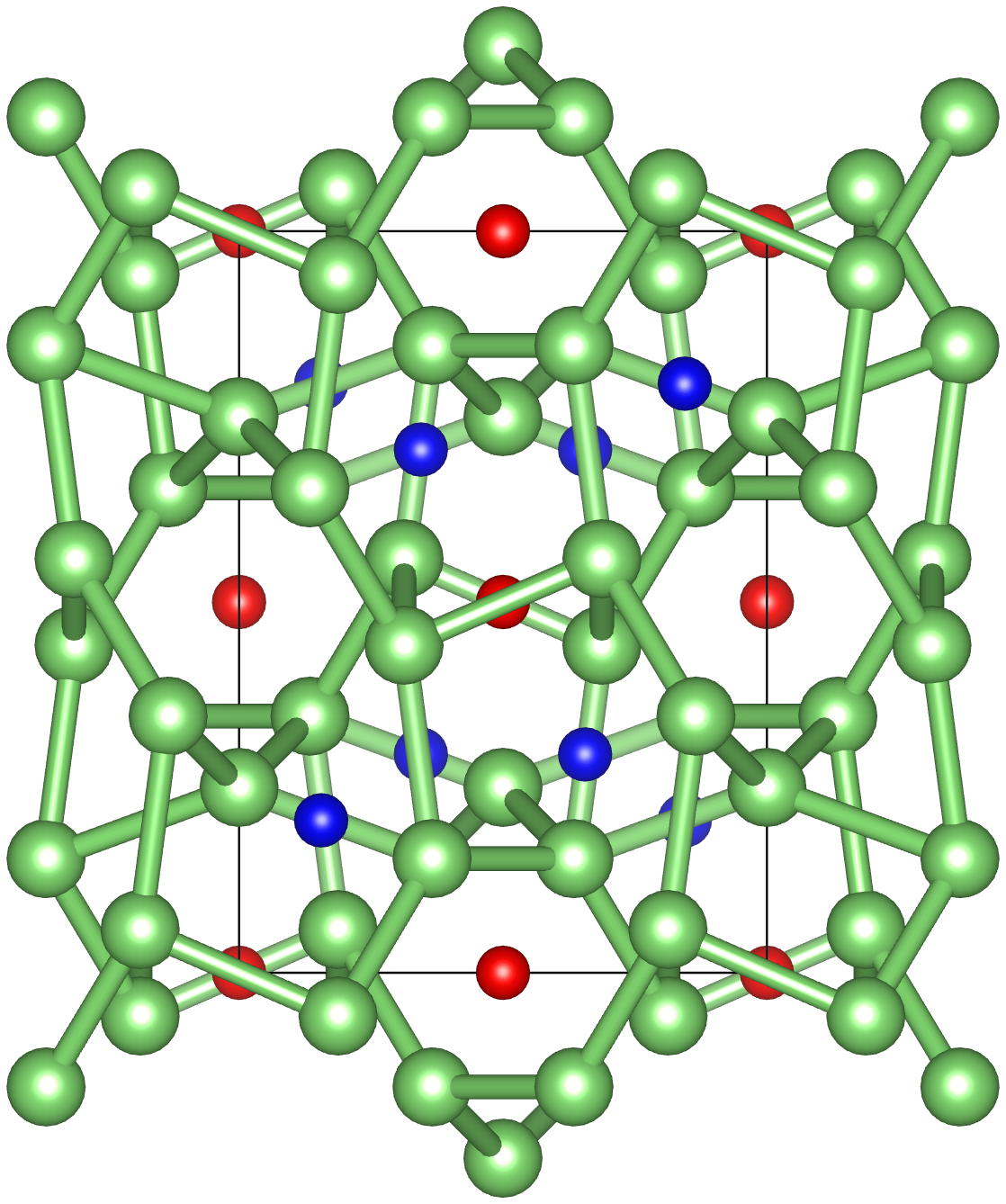}
    \caption{The cubic $I\bar{4}2m$ structure, with the two interstitial charge species shown as small red (4c) and blue (8i) spheres. Li atoms are shown in green.}
    \label{I-42m_interstitials}
\end{figure}

Fig. \ref{edos_freqs_etc} presents a summary of the key electronic properties of the structures discovered in this work, and Fig. \ref{edos_and_bandstructures} specifically compares the electronic properties of the $I\bar{4}2m$ and $Cmce$ phases. Notably, whilst the Fermi-level eDOS $\cal{N}(\epsilon_{F})$ of Li is small and relatively constant up to $\sim 16$ TPa, it nearly doubles in value upon transition to the $Cmce$ structure at $15.9$ TPa and increases almost linearly with pressure thereafter. We find that this is not because the $Cmce$ structure has a particularly large eDOS; in fact, it has $\cal{N}(\epsilon_{F})$ within $3 \%$ of the corresponding free electron value. Rather, the phases below $15.9$ TPa have unusually low $\cal{N}(\epsilon_{F})$ values. As an example, the $I\bar{4}2m$ phase at $7.50$ TPa has $\cal{N}(\epsilon_{F})$ $= 0.0270$ states eV$^{-1}$ \AA$^{-3}$, whereas the equivalent free electron value is $\cal{N}(\epsilon_{F})$ $= 0.0515$ states eV$^{-1}$ \AA$^{-3}$ (a reduction of $48 \%$). This significant eDOS reduction has been noted before in lower-pressure phases of Li \cite{Lv_2011,pickard_needs_2009}, where Li effectively becomes semiconducting. Whilst the $Fd\bar{3}m$, $Imma$, $P4_{1}32$, $I\bar{4}2m$ and $Fmm2$ structures should not be described as semiconductors, they should be considered as poor conductors relative to the $Cmce$ and higher-pressure phases. This difference is highlighted in Fig. \ref{edos}, which shows the eDOS as a function of energy for the $I\bar{4}2m$ and $Cmce$ phases; all structures below $15.9$ TPa exhibit a significant dip in the Fermi-level eDOS similar to the $I\bar{4}2m$, whereas the phases above this pressure feature no such dip and have relatively flat $\cal{N}(\epsilon_{F})$ near the Fermi level. This behaviour can be clearly linked to a sharp change in the bonding topology of Li with increasing pressure; phases below $15.9$ TPa, such as $I\bar{4}2m$, feature significant interstitial accumulations of electronic charge in spaces between atoms, forming `electrides' with localized charge that is unable to contribute to conduction. Several elements in the periodic table also form electride structures under sufficient compression \cite{Hoffman_electrides}. These electride structures exhibit electronic bandstructures with substantially flattened bands when compared to non-electride structures (compare Figs. \ref{Cmce_BS} and \ref{I-42m_BS}). Conversely, the phases above $15.9$ TPa are not electrides; they have no interstitial charge maxima, are metallically bonded, and exhibit much more dispersive bands as would be expected for a crystal under such substantial compression (again, see Fig. \ref{Cmce_BS}). We find that even up to pressures of $10$ petapascal (PPa), Li does not exhibit re-entrant electride behaviour and remains metallically bonded. Table \ref{bader_table} shows the magnitudes and Wyckoff positions of the interstitial electronic charges in the electride phases as determined using a Bader analysis \cite{bader_theory}; the $Fd\bar{3}m$, $Imma$, $P4_{1}32$, $I\bar{4}2m$ and $Fmm2$ phases all show similar interstitial accumulations of charge (see Fig. \ref{I-42m_interstitials}). The amount of localised interstitial charge per site ranges from $-0.456 \ |e|$ of charge (4e sites of the $Imma$ structure) to $-0.100 \ |e|$ of charge (8c sites of the $P4_{1}32$ structure).
\par
In the pressure range $71.9$-$103$ TPa, we find that Li adopts the incommensurate Ba-IV host-guest (HG) structure, which is assumed by several elements across the periodic table under compression \cite{HG_Aluminium,HG_Barium,HG_Bismuth_SC,HG_Calcium,HG_potassium,HG_scandium,sulfur_host_guest}. The Ba-IV HG structure comprises a `host' framework within which chains of `guest' atoms lie, and the $c$-axis periodicities of the host and guest structures are incommensurate. We calculated the enthalpy as a function of the HG ratio $\gamma = \frac{c_{H}}{c_{G}}$ at several pressures, in order to determine the ideal ratio $\gamma_{0}$ as a function of pressure (see Supplemental Material \cite{supp}). We find that the ideal HG ratio in Li decreases very slowly with increasing pressure, changing by only $\sim ~0.2 \%$ over the entire stability range of the HG phase; in contrast, the HG phase in high-pressure elemental sulfur changes its ideal ratio over the stability range by ($\sim ~12 \%$) \cite{sulfur_host_guest}. Beyond the HG phase, Li transitions to a $C2$ structure with 12 atoms in its primitive cell at $103$ TPa, and eventually to the hexagonal close-packed (hcp) structure at $116$ TPa.

\section{\label{conclusions}Conclusions}

We have performed an extensive structure search in elemental Li in the terapascal regime, uncovering 11 phase transitions beyond the previously-predicted $Fd\bar{3}m$ diamond structure. The $p$-$T$ dependence of the phase boundaries has been calculated at the level of the harmonic approximation, and the solid-liquid melting line has been determined using \textit{ab-initio} MD simulations. We have found that in the TPa regime, Li adopts the phase transition sequence $Fd\bar{3}m$ $\rightarrow$ $Imma$ $\rightarrow$ $P4_{1}32$ $\rightarrow$ $I\bar{4}2m$ $\rightarrow$ $Fmm2$ $\rightarrow$ $Cmce$ $\rightarrow$ $P\bar{6}2m$ $\rightarrow$ $C2/c \ [46]$ $\rightarrow$ $C2/c \ [22]$ $\rightarrow$ HG $\rightarrow$ $C2$ $\rightarrow$ hcp, with further transitions to structures of $P6_{3}/m$ and $P\bar{1}$ symmetry induced by temperatures of a few thousand K at pressures $25$-$37$ TPa. The $Fd\bar{3}m$, $Imma$, $P4_{1}32$, $I\bar{4}2m$ and $Fmm2$ structures are poorly conducting electrides, whereas the $Cmce$, $P\bar{6}2m$, $C2/c \ [46]$, $C2/c \ [22]$, HG, $C2$ and hcp phases are normal metals with a Fermi-level eDOS within a few percent of the corresponding free electron value.

\vspace{0.4cm}

\newpage

\section{\label{acknowledgements}Acknowledgements}

The computational resources for this project were provided by the Cambridge Service for Data Driven Discovery (CSD3), and by the UK national high performance computing service, ARCHER2, for which access was obtained via the UKCP consortium and funded by EPSRC grant refs. EP/F036884 and EP/K013564/1.

\bibliography{main.bib}

\end{document}



\title{\color{red}
Predicted Complex Lithium Phases at Terapascal Pressures \\ SUPPLEMENTAL MATERIAL
\color{black}}

\maketitle

\section{\label{pseudo}Pseudopotentials}

We used two specially constructed ultrasoft pseudopotentials for our work, generated using \verb|CASTEP|'s On-The-Fly-Generation (OTFG) program which is bundled with the code. All work up to $75$ TPa was conducted using an all-electron pseudopotential with a cutoff radius ($R_c$) of $0.60$ Bohr, which was generated using the following OTFG string:
\\\\
\texttt{Li 1|0.6|30|40|50|10U:20(qc=12)}
\\\\
This pseudopotential (hereon denoted PS1) required a plane-wave cutoff of $2.25$ keV to converge absolute energies to within $0.5$ meV (relative energies were converged to even better than this); see Supplemental Fig. \ref{06rc_pseudo_cutoff}.
\\\\
All work above $75$ TPa was conducted using an all-electron pseudopotential with $R_{c} = 0.40$ Bohr, which was generated using the following OTFG string:
\\\\
\texttt{Li 1|0.4|80|90|100|10U:20(qc=16.5)}
\\\\
This pseudopotential (hereon denoted PS2) required a plane-wave cutoff of $4.50$ keV to converge absolute energies to within $0.5$ meV (relative energies were converged to even better than this); see Supplemental Fig. \ref{04rc_pseudo_cutoff}.

\begin{figure}[h!]

    \begin{subfigure}{0.500\textwidth}
        \centering
        \hspace{-0.75cm}
        \includegraphics[width=0.850\textwidth]{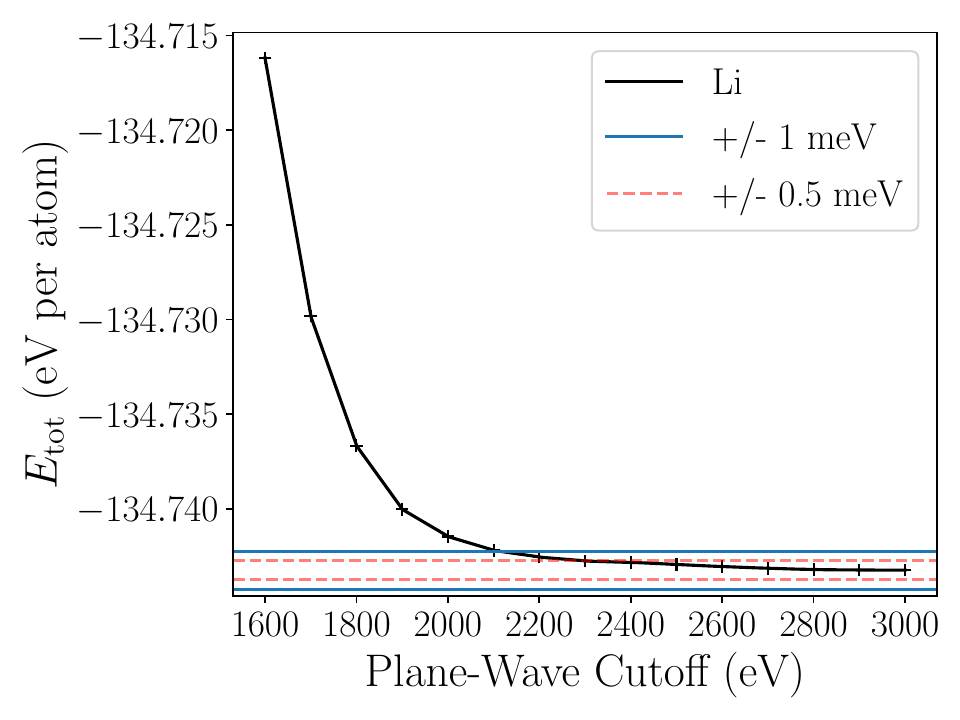}
        \caption{PS1 convergence ($P\bar{6}2m$ structure at 30.0 TPa).}
        \label{06rc_pseudo_cutoff}
    \end{subfigure}

    \vspace{0.5cm}
    
    \begin{subfigure}{0.500\textwidth}
        \centering
        \hspace{-0.75cm}
        \includegraphics[width=0.850\textwidth]{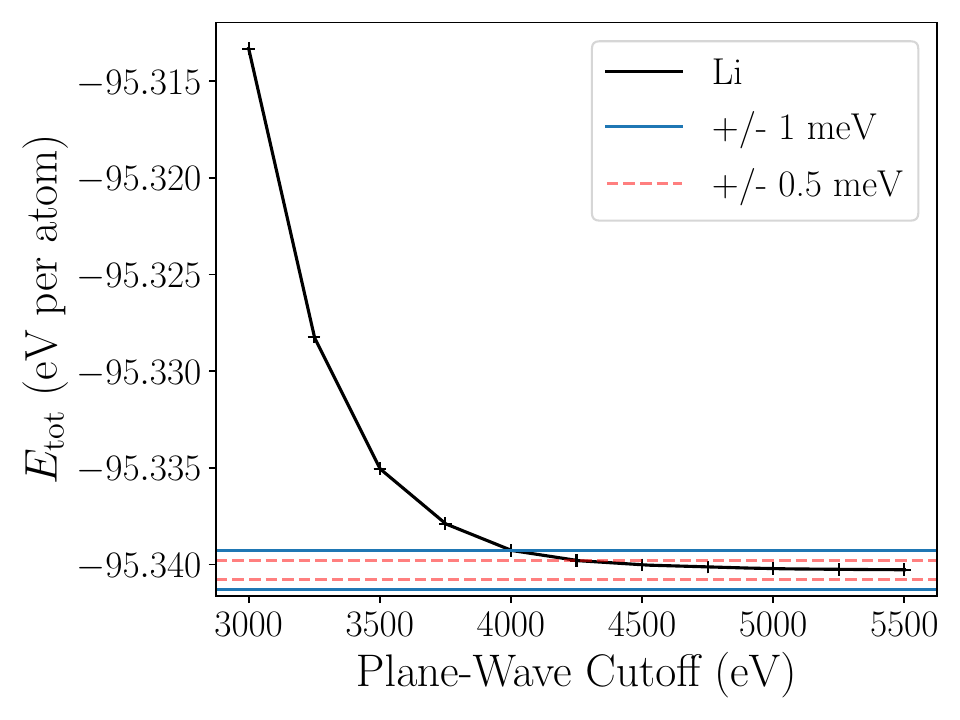}
        \caption{PS2 convergence ($C2/c \ [22]$ structure at 65.0 TPa).}
        \label{04rc_pseudo_cutoff}
    \end{subfigure}
    
    \caption{Convergence of the total energy with plane-wave cutoff for the two pseudopotentials used in this work, using some representative structures.}
    \label{cutoff_convergence}
    
\end{figure}

\clearpage

\section{\label{convergence}k-point Sampling Convergence}

\begin{figure}[h!]
    \centering
    \hspace{-0.4cm}
    \includegraphics[width=0.475\textwidth]{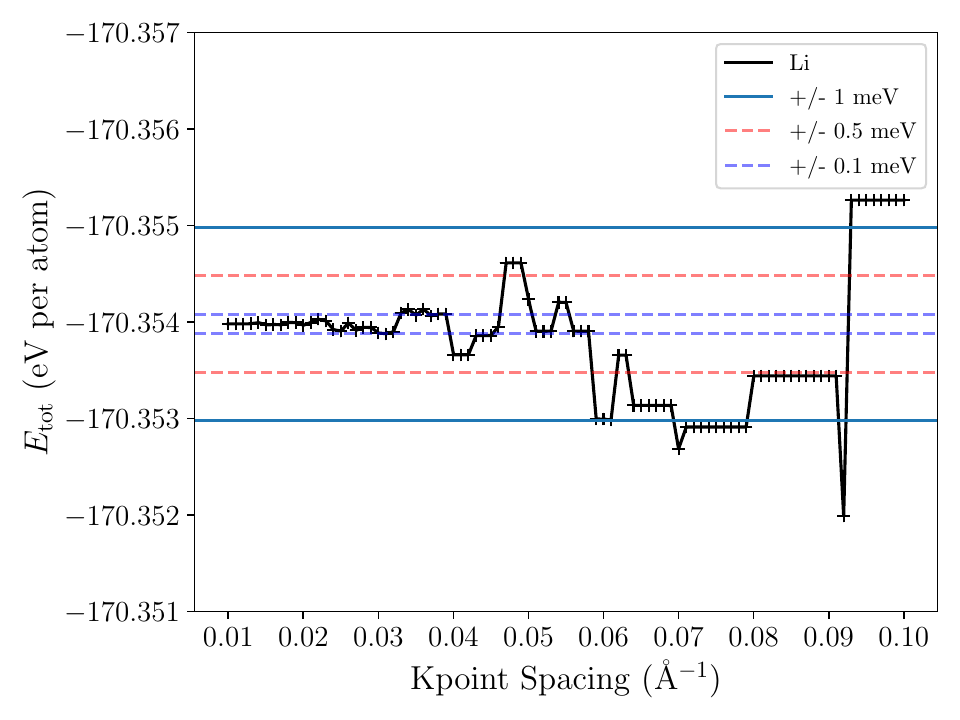}
    \caption{Convergence of total energy of $I\bar{4}2m$ at 8.0 TPa with respect to \textbf{k}-point sampling density.}
    \label{kpoints_convergence}
\end{figure}

Supplemental Fig. \ref{kpoints_convergence} shows the convergence of the total energy of $I\bar{4}2m$ lithium as a function of \textbf{k}-point sampling density at $8.0$ TPa. It can be seen that even with a relatively coarse \textbf{k}-point spacing of $0.060$ \AA $^{-1}$, total energies are converged to better than $\pm 1$ meV. Convergence to better than $\pm 0.1$ meV is achieved at spacings of $0.035$ \AA $^{-1}$ or finer. As stated in the main manuscript, we used a \textbf{k}-point spacing of of $0.025$ \AA $^{-1}$.

\clearpage

\section{\label{SL_curves} Static Lattice Enthalpies}

\begin{figure}[h!]

    \begin{subfigure}{0.500\textwidth}
        \centering
        \hspace{-0.75cm}
        \includegraphics[width=0.850\textwidth]{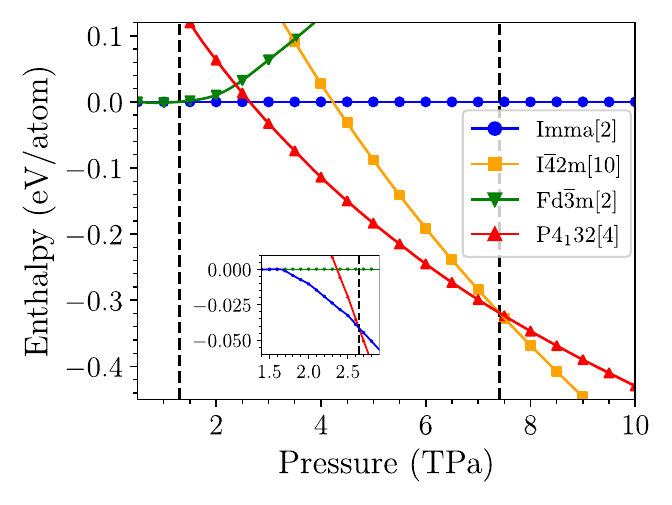}
        \caption{SL low.}
        \label{SL_low}
    \end{subfigure}

    \vspace{0.5cm}
    
    \begin{subfigure}{0.500\textwidth}
        \centering
        \hspace{-0.75cm}
        \includegraphics[width=0.850\textwidth]{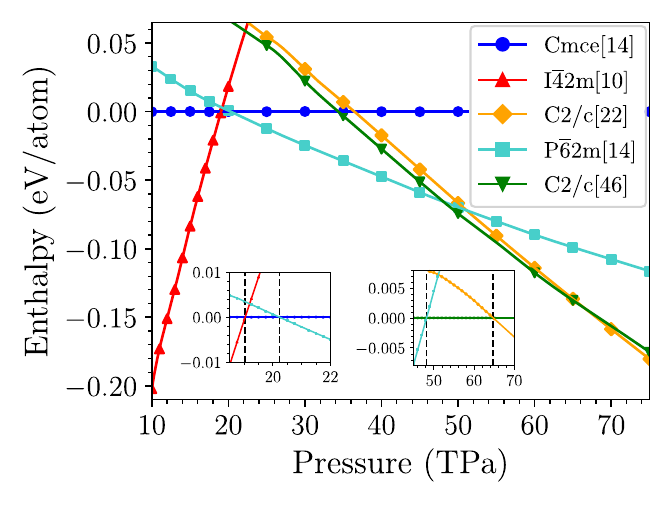}
        \caption{SL medium.}
        \label{SL_medium}
    \end{subfigure}

    \vspace{0.5cm}

    \begin{subfigure}{0.500\textwidth}
        \centering
        \hspace{-0.75cm}
        \includegraphics[width=0.850\textwidth]{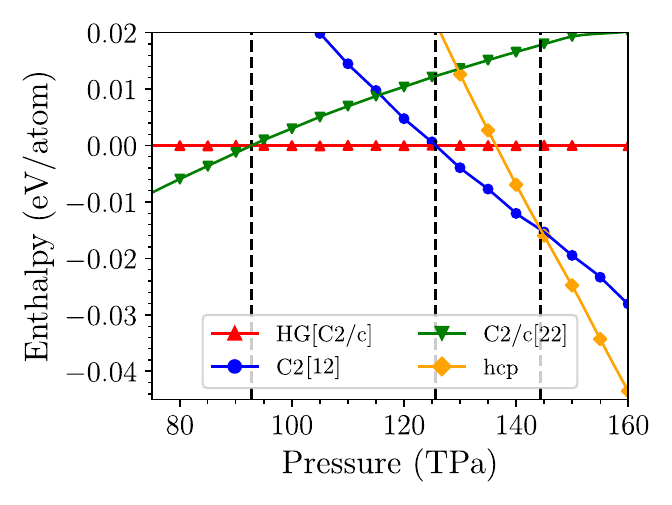}
        \caption{SL high.}
        \label{SL_high}
    \end{subfigure}
    
    \caption{Static-lattice PBE-DFT enthalpies as a function of pressure. The numbers in square brackets denote the number of atoms in the primitive cell. Dashed vertical lines mark the static-lattice phase transitions.}
    \label{SL_enthalpies}
    
\end{figure}

\clearpage

\section{\label{HG_ratio} Pressure Dependence of Ba-IV Host-Guest Ideal Ratio}

\begin{figure}[h!]
    \centering
    \hspace{-0.4cm}
    \includegraphics[width=0.475\textwidth]{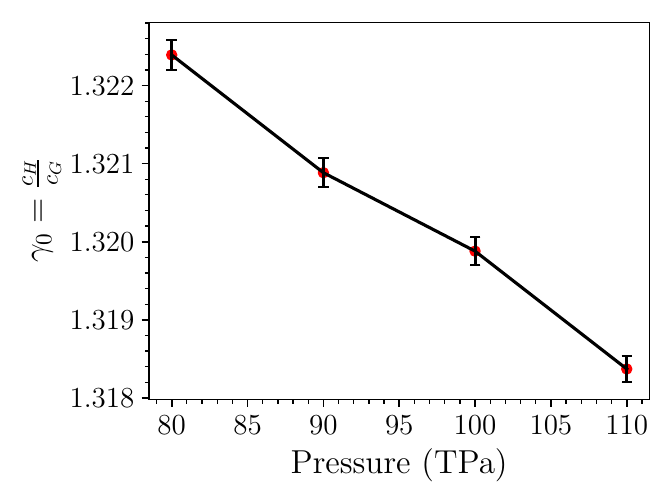}
    \caption{The ideal HG ratio $\gamma_{0}$ as a function of pressure.}
    \label{ratio_vs_pressure}
\end{figure}

Supplemental Fig. \ref{ratio_vs_pressure} shows the ideal host-guest (HG) ratio $\gamma_{0}$ for the incommensurate Ba-IV-type phase as a function of pressure.

\begin{figure}[h!]
    \centering
    \hspace{-0.4cm}
    \includegraphics[width=0.475\textwidth]{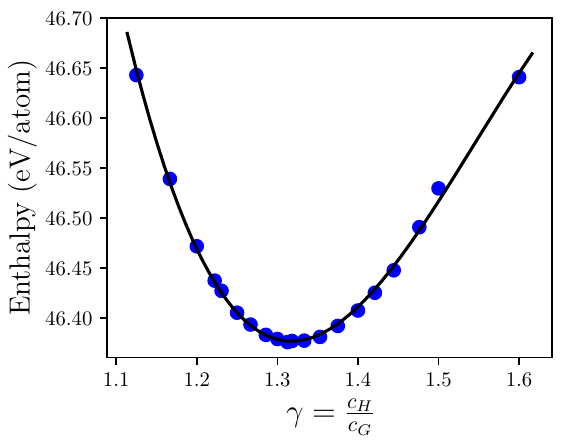}
    \caption{Enthalpy as a function of HG ratio $\gamma$, at $90$ TPa. Each marker represents a particular commensurate HG approximant of the Ba-IV type. For this pressure, the ideal ratio is $\gamma_{0} = 1.3209$.}
    \label{enthalpy_vs_ratio}
\end{figure}

For a given pressure, the ideal HG ratio is determined by calculating the total enthalpy $H$ for several commensurate Ba-IV approximants of differing HG ratio, and then fitting the resulting data to a curve. We found that near the ideal ratio, a cubic fit of the form:

\begin{equation}
    H(\gamma) = a + b(\gamma-\gamma_{0})^{2} + c(\gamma-\gamma_{0})^{3}
\end{equation}

where $(\gamma_{0},a,b,c)$ are the parameters to be fitted, fit the data very well, typically leading to an error in the ideal ratio of $\sim 2 \times 10^{-4}$ at each pressure. Fitting functions with higher powers were investigated, but they offered a negligable improvement on the error.
\\\\
Once the data has been fit, one can obtain the ideal ratio $\gamma_{0}$ at that given pressure. Supplemental Fig. \ref{enthalpy_vs_ratio} shows an example of such a fit at $90$ TPa; the ideal ratio can be read off here as $\gamma_{0} = 1.3209$. When multiple such fits are carried out at different pressures, $\gamma_{0}$ can be obtained as a function of pressure (see Supplemental Fig. \ref{ratio_vs_pressure} above).

\clearpage

\section{\label{finite_T} Finite Temperature Calculations}

Fig. 1 in the main manuscript is a $p$-$T$ phase diagram at the level of the vibrational harmonic approximation. This section details how these finite-temperature calculations were carried out.
\\\\
All phonon calculations were done with the \verb|CAESAR| Harmonic Calculation Library, with the \verb|CASTEP| DFT code as the underlying force calculator. For each structure, phonons were calculated on a coarse $8 \times 8 \times 8$ \textbf{q}-point grid using a finite-differences displacement of $0.1$ Bohr. Once calculated on the coarse grid, the phonon frequencies were interpolated onto a finer $40 \times 40 \times 40$ \textbf{q}-point grid, upon which vibrational observables were calculated. This procedure was sufficient to converge total vibrational energies to better than $\pm 0.5$ meV. 
\\\\
Once the phonons had been calculated according to the above procedure, the total Helmholtz free energy at finite temperature $T$ was calculated via:

    \begin{equation}
	    F(T,V)  =  U_{elec}(T,V)  +  U_{ph}(T,V)  -  TS_{elec}  -  TS_{ph}
    \end{equation}

Where `elec’ and `ph’ denote electronic and phononic contributions to $F(T,V)$, respectively. $F_{elec}$ at finite temperatures was calculated by thermally occupying the Kohn-Sham orbitals using the Fermi-Dirac occupation function. We note however, that for all phases and temperatures discussed in this work, $\frac{T}{T_{Fermi}} << 1$. As such, $TS_{elec} << TS_{ph}$, and the finite-temperature contributions to the free energy are dominated by the vibrational terms.
\\\\
The phonon total energy $U_{ph}$ and vibrational entropy $S_{ph}$ were calculated using the harmonic phonon DOS and thermally occupying states according to the Bose-Einstein distribution.
\\\\
$F(T,V)$ was then fitted to a univariate spline using the \verb|scipy| python package. It is commonplace to instead use an equation of state (EOS) such as the Vinet EOS or Birch-Murnaghan EOS rather than a spline fit - however, at such enormous pressures, these EOSs are unreliable, and were found to give a poor fit to the data.
\\\\
Once $F(T,V)$ had been fitted, the pressure at a given temperature was then found from the (negative) volume derivative of $F(T,V)$, and thus the total Gibbs free energy ($G = F + pV$) at pressure $p$ can be calculated as:

    \begin{equation}
	    G(T,p) = \bigg[ F(T,V) \underbrace{\ - \ \frac{\partial F(T,V)}{\partial V}}_{p} \cdot V \bigg]_{\mathrm{min}(V)}
    \end{equation}

This approach to calculating $G(T,p)$ includes (harmonic) phonon corrections to the pressure.
\\\\
As an example, the phonon dispersion of the $Cmce$ structure at $17.0$ TPa, as obtained using \verb|CAESAR| + \verb|CASTEP|, is shown in Supplemental Fig. \ref{example_ph_dispersion} below.

\begin{figure}[h!]
    \centering
    \includegraphics[width=0.450\textwidth]{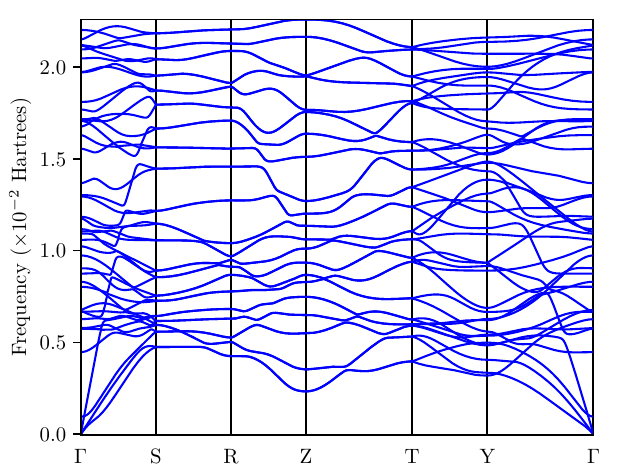}
    \caption{Phonon dispersion of $Cmce$ Li at $17.0$ TPa, along selected high symmetry paths.}
    \label{example_ph_dispersion}
\end{figure}

\clearpage

\section{\label{ab_initio_MD} \textit{Ab-Initio} Molecular Dynamics Simulations for Melting}

We performed \textit{ab-inito} molecular dynamics simulations to calculate the melting temperature of Li as a function of pressure. Melting was calculated by finding the temperature (in an NPT ensemble) at which the root-mean-squared displacement for all atoms diverged (`heat until melting' approach).
\\\\
The simulations were run in an NPT ensemble using the \verb|CASTEP| code, using a timestep of $0.5$ fs, for a total of 25,000 timesteps (i.e. 20 ps). Temperature was controlled using a Nosé-Hoover thermostat, and pressure controlled with an Andersen-Hoover barostat. The particular $(p,T)$ pairs that were explicitly simulated can be found as red crosses on Fig. 1 in the main manuscript.
\\\\
Simulations were converged with respect to supercell size, and for reasons of computational expense, the smallest such converged supercell was used in each simulation. Cells were chosen to be shaped as cubic as possible. The resulting supercells for each simulation are shown in supplemental Table \ref{supercells_table}.

\renewcommand{\arraystretch}{1}
\begin{table}[h!]
\begin{tabular}{c c c}
\hline\hline
\centering
Structure & Supercell & $N_{atoms}$ \\ \hline

$Fd\bar{3}m$  & 3 $\times$ 3 $\times$ 3 & 216 \\
$Imma$        & 4 $\times$ 4 $\times$ 3 & 192 \\
$P4_{1}32$    & 4 $\times$ 4 $\times$ 4 & 256 \\
$I\bar{4}2m$  & 2 $\times$ 2 $\times$ 2 & 160 \\
$P\bar{6}2m$  & 3 $\times$ 3 $\times$ 2 & 252 \\
$C2/c \ [46]$ & 2 $\times$ 4 $\times$ 2 & 1472 \\
$C2/c \ [22]$ & 2 $\times$ 3 $\times$ 2 & 528 \\
HG            & 3 $\times$ 3 $\times$ 2 & 576 \\
$C2$          & 2 $\times$ 4 $\times$ 2 & 384 \\
hcp           & 7 $\times$ 7 $\times$ 4 & 392 \\

\hline\hline
\end{tabular}
\captionof{table}{Supercells (with respect to the conventional cell) used for the \textit{ab-initio} MD simulations. For some structures, as in the main manuscript, numbers in square brackets denote the number of atoms in the primitive cell.}
\label{supercells_table}
\end{table}

\addtolength{\tabcolsep}{+2.5pt} 
\renewcommand{\arraystretch}{1.25}
\begin{table*}[h!]
\begin{tabular}{l l l l}
\hline\hline
\centering
Structure     & Pressure (TPa) & Structural Parameters (\AA)        & Wyckoff Positions                  \\  \hline
$Fd\bar{3}m$  & 0.50           & $a=2.8470$                         & 8a                                 \\  \hline
$Imma$        & 2.00           & $a=1.4316 , b=1.6858 , c=2.5087$   & 4e $(z=0.5992)$                    \\  \hline
$P4_{1}32$    & 5.00           & $a=1.5681$                         & 4a                                 \\  \hline
$I\bar{4}2m$  & 7.50           & $a=2.2387 , c=3.1398$              & 8i $(x=0.1337 , z=0.1533)$         \\
              &                &                                    & 8i $(x=0.6879 , z=0.9412)$         \\
              &                &                                    & 4d                                 \\  \hline
$Fmm2$        & 13.0           & $a=2.8717 , b=2.9073 , c=2.8842$   & 8d $(x=0.3202,z=0.0078)$           \\
              &                &                                    & 8c $(y=0.1326,z=0.0996)$           \\
              &                &                                    & 8b $(z=0.1940)$                    \\
              &                &                                    & 8d $(x=0.3624,z=0.2979)$           \\
              &                &                                    & 8c $(y=0.1947,z=0.3847)$           \\  \hline
$Cmce$        & 17.5           & $a=3.2008 , b=1.6134 , c=2.8022$   & 4b                                 \\
              &                &                                    & 16g $(x=0.1918,y=0.2382,z=0.1267)$ \\
              &                &                                    & 8f $(y=0.5915,z=0.6613)$           \\  \hline
$P\bar{6}2m$  & 30.0           & $a=1.5727 , c=2.5864$              & 3f $(x=0.2848)$                    \\
              &                &                                    & 6i $(x=0.5800,z=0.2185)$           \\
              &                &                                    & 2e $(z=0.3535)$                    \\
              &                &                                    & 3g $(z=0.4485)$                    \\  \hline
$C2/c \ [46]$ & 50.0           & $a=5.1348 , b=1.4297 , c=3.8179$   & 4e $(y=0.6780)$                    \\
              &                & $\beta=90.148 \degree$             & 8f $(x=0.3864,y=0.4338,z=0.0338)$  \\
              &                &                                    & 8f $(x=0.7965,y=0.3517,z=0.0467)$  \\
              &                &                                    & 8f $(x=0.9327,y=0.4551,z=0.0575)$  \\
              &                &                                    & 8f $(x=0.5230,y=0.3275,z=0.0769)$  \\
              &                &                                    & 8f $(x=0.2636,y=0.3130,z=0.0941)$  \\
              &                &                                    & 8f $(x=0.1342,y=0.4626,z=0.1052)$  \\
              &                &                                    & 8f $(x=0.6638,y=0.4435,z=0.1321)$  \\
              &                &                                    & 8f $(x=0.0244,y=0.2496,z=0.1585)$  \\
              &                &                                    & 8f $(x=0.9015,y=0.0281,z=0.2102)$  \\
              &                &                                    & 8f $(x=0.3624,y=0.0446,z=0.2109)$  \\  
              &                &                                    & 8f $(x=0.7652,y=0.1758,z=0.2214)$  \\ \hline
$C2/c \ [22]$ & 65.0           & $a=3.8421 , b=1.3881 , c=2.4525$   & 4e $(y=0.6852)$                    \\
              &                & $\beta=116.372 \degree$            & 8f $(x=0.3351,y=0.3177,z=0.0143)$  \\
              &                &                                    & 8f $(x=0.6015,y=0.0478,z=0.1065)$  \\
              &                &                                    & 8f $(x=0.8371,y=0.3093,z=0.1098)$  \\  
              &                &                                    & 8f $(x=0.0086,y=0.1298,z=0.1161)$  \\
              &                &                                    & 8f $(x=0.7036,y=0.4342,z=0.2380)$  \\ \hline
HG ($\gamma=\frac{4}{3}$) & 90.0 & $a=1.8410 , b=1.7673 , c=2.3886$ & 8f $(x=0.7319,y=0.1270,z=0.0845)$  \\ 
              &                & $\beta=112.028 \degree$            & 8f $(x=0.3769,y=0.1469,z=0.1107)$  \\
              &                &                                    & 8f $(x=0.5724,y=0.4574,z=0.1390)$  \\ 
              &                &                                    & 8f $(x=0.0995,y=0.3284,z=0.1670)$  \\ \hline
$C2$          & 105            & $a=2.2345 , b=1.2028 , c=1.8550$   & 2a ($y=0.2873$)                    \\
              &                & $\beta=94.1705728 \degree$         & 2b ($y=0.2010$)                    \\
              &                &                                    & 4c $(x=0.2280,y=0.4541,z=0.0884)$  \\
              &                &                                    & 4c $(x=0.9482,y=0.2859,z=0.1652)$  \\
              &                &                                    & 4c $(x=0.6800,y=0.4907,z=0.2461)$  \\ 
              &                &                                    & 4c $(x=0.4380,y=0.2312,z=0.3333)$  \\
              &                &                                    & 4c $(x=0.7153,y=0.0439,z=0.4198)$  \\ \hline
hcp           & 120            & $a=0.6543 , c=1.0415$              & 2c                                 \\

\hline\hline
\end{tabular}
\captionof{table}{Structural parameters for the structures discussed in this work, at selected pressures.}
\label{structure_table}
\addtolength{\tabcolsep}{2.5pt} 
\end{table*}